\documentclass[12pt]{article}
\usepackage{amsmath,amssymb,url,epsfig,color,hyperref}

\def\m@th{\mathsurround=0pt }
\def\leftrightarrowfill{$\m@th \mathord\leftarrow \mkern-6mu
        \cleaders\hbox{$\mkern-2mu \mathord- \mkern-2mu$}\hfill
        \mkern-6mu \mathord\rightarrow$}

\def\overleftrightarrow#1{\vbox{\ialign{##\crcr
        \leftrightarrowfill\crcr\noalign{\kern-1pt\nointerlineskip}
        $\hfil\displaystyle{#1}\hfil$\crcr}}}

\newcommand{\gev}{\hbox{\rm\,GeV}}
\newcommand{\mev}{\hbox{\rm\,MeV}}

\newcommand{\dr}{\mbox{$\Delta r$}}
\newcommand{\drcarw}{\mbox{$\Delta \hat{r}_{W}$}}
\newcommand{\drcarwu}{\mbox{$\Delta \hat{r}^{(1)}_{W}$}}
\newcommand{\drcarwd}{\mbox{$\Delta \hat{r}^{(2)}_{W}$}}

\newcommand{\bms}{\mbox{$\overline{MS}$}}
\newcommand{\ms}{\mbox{$\overline{\scriptstyle MS}$}}
\newcommand{\sincur}{\mbox{$\sin^{2}\!\hat{\theta}_{W}$}}
\newcommand{\sinw}{\mbox{$\sin^{2}\!\theta_{W}$}}
\newcommand{\scs}{\mbox{$\hat{s}^{2}$}}
\newcommand{\sq}{\mbox{$s^{2}$}}
\newcommand{\cq}{\mbox{$c^{2}$}}
\newcommand{\ccs}{\mbox{$\hat{c}^{2}$}}
\newcommand{\alc}{\mbox{$\hat{\alpha}$}}
\newcommand{\agg}{A_{\gamma\gamma}}
\newcommand{\pigg}{\Pi_{\gamma\gamma}}
\newcommand{\piggp}[1]{\mbox{$\Pi^{\mbox{\scriptsize{(#1)}}}_{\gamma\gamma}$}}

\newcommand{\aww}{A_{\scriptscriptstyle WW}}
\newcommand{\azz}{A_{\scriptscriptstyle ZZ}}
\newcommand{\agz}{A_{\scriptscriptstyle \gamma Z}}
\newcommand{\Dah}{\Delta \alpha^{(5)}_{\rm had}(\mz^2)}
\newcommand{\rhoh}{\mbox{$\hat{\rho}$}}

\newcommand{\newc}{\newcommand}

\newc{\gsim}{\lower.7ex\hbox{$\;\stackrel{\textstyle>}{\sim}\;$}}
\newc{\lsim}{\lower.7ex\hbox{$\;\stackrel{\textstyle<}{\sim}\;$}}
\newc{\mt}{M_t}
\newc{\mb}{m_b}
\newcommand{\mw}{m_{\scriptscriptstyle W}}

\newcommand{\Mw}{M_{\scriptscriptstyle W}}
\newcommand{\Mz}{M_{\scriptscriptstyle Z}}
\newcommand{\Gz}{\Gamma_{\scriptscriptstyle Z}}
\newcommand{\mh}{m_{\scriptscriptstyle H}}
\newcommand{\mz}{m_{\scriptscriptstyle Z}}
\newcommand{\mwo}{m_{\scriptscriptstyle W_{0}}}
\newcommand{\Mzo}{M_{\scriptscriptstyle Z_{0}}}
\newcommand{\mzo}{m_{\scriptscriptstyle Z_{0}}}
\newcommand{\as}{\alpha_s}

\newc{\alphasmz}{\alpha_s(\mz)}

\newcommand{\mhw}{\zeta_{\scriptscriptstyle W}}
\newcommand{\mhz}{\zeta_{\scriptscriptstyle Z}}
\newcommand{\mtw}{t_{\scriptscriptstyle W}}
\newcommand{\mtz}{t_{\scriptscriptstyle Z}}
\newcommand{\mbw}{b_{\scriptscriptstyle W}}
\newcommand{\mbz}{b_{\scriptscriptstyle Z}}
\newcommand{\lnb}{\mbox{$\overline{\mbox{ln}}$}}

\newcounter{mysubequation}[equation]

\def\be{\begin{equation}}
\def\ee{\end{equation}}
\def\bea{\begin{eqnarray}}
\def\eea{\end{eqnarray}}

\def\slashchar#1{\setbox0=\hbox{$#1$}           
   \dimen0=\wd0                                 
   \setbox1=\hbox{/} \dimen1=\wd1               
   \ifdim\dimen0>\dimen1                        
      \rlap{\hbox to \dimen0{\hfil/\hfil}}      
      #1                                        
   \else                                        
      \rlap{\hbox to \dimen1{\hfil$#1$\hfil}}   
      /                                         
   \fi}                                         %
\catcode`@=11
\long\def\@caption#1[#2]#3{\par\addcontentsline{\csname
  ext@#1\endcsname}{#1}{\protect\numberline{\csname
  the#1\endcsname}{\ignorespaces #2}}\begingroup
    \small
    \@parboxrestore
    \@makecaption{\csname fnum@#1\endcsname}{\ignorespaces #3}\par
  \endgroup}
\catcode`@=12

\usepackage{multicol}
\usepackage{color}
\definecolor{rosso}{cmyk}{0,1,1,0.4}
\definecolor{rossos}{cmyk}{0,1,1,1}
\definecolor{rossoc}{cmyk}{0,1,1,0.2}
\definecolor{blu}{cmyk}{1,1,0,0.3}
\definecolor{blus}{cmyk}{1,1,0,0.6}
\definecolor{bluc}{cmyk}{1,1,0,0.1}
\definecolor{verde}{cmyk}{0.92,0,0.59,0.25}
\definecolor{verdec}{cmyk}{0.92,0,0.59,0.15}
\definecolor{verdes}{cmyk}{0.92,0,0.59,0.4}
\definecolor{grigio}{cmyk}{0,0,0,0.07}
\definecolor{rosa}{cmyk}{0,0.1,0.1,0.02}
\definecolor{rosino}{cmyk}{0,0.05,0.05,0.02}
\definecolor{rosas}{cmyk}{0,0.3,0.25,0.05}
\definecolor{celeste}{cmyk}{0.1,0,0,0.02}
\definecolor{giallino}{cmyk}{0,0,0.4,0.02}
\definecolor{rosso}{cmyk}{0,1,1,0.4}
\definecolor{rossos}{cmyk}{0,1,1,0.55}
\definecolor{rossoc}{cmyk}{0,1,1,0.2}
\definecolor{blu}{cmyk}{1,1,0,0.3}
\definecolor{bluc}{cmyk}{1,1,0,0.1}
\definecolor{blucc}{cmyk}{0.7,0.5,0,0}
\definecolor{viola}{cmyk}{0,1,0,0.6}
\definecolor{viola2}{cmyk}{0,1,0.2,0.6}
\definecolor{verde}{cmyk}{0.92,0,0.59,0.25}
\definecolor{verdec}{cmyk}{0.92,0,0.59,0.15}
\definecolor{verdes}{cmyk}{0.92,0,0.59,0.4}
\definecolor{verdino}{cmyk}{0.12,0,0.09,0.05}
\definecolor{giallo}{cmyk}{0,0,1,0}
\definecolor{gialloverde}{cmyk}{0.44,0,0.74,0}

\newenvironment{appendletterA}
 {
  \setcounter{section}{0}
  \setcounter{equation}{0}
  
 }{
 }

\begin{document}\hfill

{\flushright{
        \begin{minipage}{3.0cm}
          RM3-TH/14-19
        \end{minipage}        }

}


\color{black}
\vspace{1cm}
\begin{center}
{\Large \bf \color{magenta} The $\mw-\mz$ interdependence in the Standard Model:
a new scrutiny}

\bigskip\color{black}\vspace{0.6cm}{
{\large\bf  Giuseppe~Degrassi$^{a}$, Paolo~Gambino$^{b}$, 
Pier Paolo~Giardino$^{a}$}
} \\[7mm]
{\it  (a) Dipartimento di Matematica e Fisica, Universit{\`a} di Roma Tre and \\
 INFN, sezione di Roma Tre, I-00146 Rome, Italy}\\[1mm]
{\it (b) Dipartimento di Fisica, Universit{\`a} di Torino and \\
INFN, sezione di Torino, I-10125 Turin, Italy}\\[1mm]
\end{center}
\bigskip
\bigskip
\bigskip
\vspace{1cm}

\centerline{\large\bf Abstract}
\begin{quote}
The  $\mw-\mz$ interdependence in the Standard Model is studied at 
$O(\alpha^2) $ 
in the $\bms$ scheme. The relevant radiative parameters, 
$\Delta \alc (\mu),\, \drcarw,\, \rhoh$ are computed 
at the full two-loop level
augmented by higher-order QCD contributions  and by 
resummation of reducible contributions. 
 We obtain $\mw = 80.357 \pm
0.009 \pm 0.003$ GeV where the errors refer to the parametric and theoretical uncertainties, respectively.
A comparison with the known result in the 
On-Shell scheme gives a difference of  $ \approx 6$ MeV. 
As a byproduct
of our calculation we also obtain the \bms\ electromagnetic coupling and the
Weinberg angle at the top mass scale, $\alc (\mt) =  (127.73)^{-1} \pm
0.0000003$ and $\sincur (\mt) = 0.23462 \pm  0.00012$.
\end{quote}
\thispagestyle{empty}
\newpage
\section{Introduction}
The discovery of a new scalar resonance
 with mass around 125 GeV and properties compatible with those of the
 Standard Model (SM) Higgs boson at the Large Hadron Collider (LHC)
 \cite{higgsdiscovery} has completed the search for the particles
 foreseen in the SM.  The first run of the LHC has delivered two
 important messages: i) no signal of physics beyond the SM (BSM) was
 observed. ii) The Higgs boson was found exactly in the mass range
 110--160 GeV predicted by the SM.  This indicates that BSM physics,
 if it exists, is likely to be at a high scale, and possibly out of
 the reach of direct LHC searches, in which case BSM physics could be
 constrained only indirectly using high precision measurements.  To
 this end, the precision of the SM predictions should match the
 experimental one and the theoretical uncertainties should be reliably
 estimated.


Among the various precision observables the  $W$ boson mass, 
$\mw$, has always played a very important role. Historically, the inclusion
of the radiative corrections in the prediction of 
$\mw$  in the SM from the 
electromagnetic coupling, $\alpha$, the Fermi 
constant, $G_\mu$, and the Weinberg angle, $\theta_W$, as extracted form deep 
inelastic neutrino scattering \cite{Si81}, was the main motivation to develop 
the 
On-Shell (OS) renormalization scheme \cite{Si80}. In the OS scheme 
$\theta_W$ is defined
in terms of the pole masses of the $W$ and $Z$ bosons, $\sinw \equiv
\sq = 1- \mw^2/\mz^2$, and the tree level relation between $G_\mu$ and
$\mw$ is corrected by the radiative parameter $\dr$ via
\be
 \frac{G_\mu}{\sqrt{2}}  = \frac{\pi \alpha}{2 \mw^2 \sq} \left[ 1 +
\dr \right]
\label{dr}
\ee
that gives rise to an $\mw-\mz$ interdependence expressed by
\be
        \mw^2 = \frac{ \mz^2}{2} \left\{
         1 + \left[ 1 - \frac{4 A^2}{\mz^2} (1+\dr) \right]^{1/2}
                            \right\}~ ,
\label{osmwmz}
\ee
where $A= (\pi \alpha/(\sqrt{2} G_\mu))^{1/2} = 37.2804(3)$ GeV.

Since the pioneering one-loop computation of $\dr$ reported in
Ref.\cite{Si80} many studies have been devoted to the calculation of
higher-order (two or more loops) effects in $\dr$. First, the
higher-order contribution related to the iteration of the large
one-loop term of ${\cal O}(\alpha \ln(\mz/m_f))$, where $m_f$ is a
generic light fermion mass was investigated \cite{Si84}.  Then, strong
and electroweak (EW) corrections to the one-loop $\delta \rho$
contribution, and in particular the effects proportional to powers of
the top mass, were investigated in detail for vanishing bottom mass.
The ${\cal O}(\alpha \as)$ contribution to $\delta \rho$ was obtained
in Ref.\,\cite{QCD2lrho}, and later the three-loop calculation ${\cal
  O}(\alpha \as^2)$ was also accomplished \cite{QCD3lrho,QCD3lrhon}.
Concerning the EW corrections, the leading two-loop contribution
${\cal O}(\alpha^2 \mt^4/\mw^4)$ to $\delta \rho$ was first obtained
in the large top-mass limit \cite{Hoo}, neglecting all the other
masses including the Higgs mass, and then in the so-called gaugeless
limit of the SM, i.e. in the limit $g, \, g' \to 0$ where $g$ ($g'$) is the $SU(2)$ ($U(1)_Y$) gauge coupling
\cite{bar}. The
incorporation of these effects in $\dr$ was addressed in
Ref.\,\cite{Con}.  Needless to say, these calculations were
instrumental in the successful prediction of the top mass before its
actual discovery.  The two-loop knowledge of $\dr$ was later improved
with the evaluation of the next-to-leading effects in the heavy top
expansion, namely the ${\cal O}(\alpha^2 \mt^2/\mw^2)$ contributions
\cite{DGV, DGS}. The latter turned out to be comparatively large and
allowed for a drastic reduction of the scheme dependence.  Leading
three-loop effects related to $\delta \rho$, in particular the ${\cal
  O}(\alpha^3 \mt^6/\mw^6)$ and ${\cal O}(\alpha^2 \as \mt^4/\mw^4)$
contributions, were also investigated \cite{FKSV}.

The complete calculation of $\dr$ at the two-loop level was accomplished
in several steps. First, the $ {\cal O}(\alpha \as)$ corrections were obtained
from the full QCD corrections to the gauge bosons 
self-energies \cite{QCD2l}. Then the two-loop fermionic contribution, i.e.\
two-loop diagrams with at least one  closed fermion loop,  was derived
\cite{FHWW, AC}, and finally the purely bosonic contribution was also obtained
\cite{ACOV}, completing the two-loop computation of $\dr$. The 
prediction of $\mw$  from eq.\,(\ref{osmwmz})  at the two-loop accuracy, 
including also  known three-loop effects, was summarized in Ref.~\cite{ACFW} by 
a simple formula that parameterizes the result in term of
the relevant input quantities, used 
by several groups in their fits to the EW precision observables
\cite{fitpro,Gfitter, CFMS}. Ref.~\cite{ACFW} also  estimated the uncertainty due to 
unknown 
higher-order effects, $\delta \mw^{th} \approx 4$ MeV, from the size of the computed three-loop
corrections.

The present experimental world average $\mw^{exp} = 80.385 \pm 0.015$
GeV agrees well with the indirect determination of $\mw$ via a full fit to EW
precision observables (except $\mw$): Ref.~\cite{Gfitter} reports
$\mw^{fit} = 80.359 \pm 0.011$, consistent with the result of
Ref.~\cite{CFMS}, $\mw^{fit} = 80.362 \pm 0.007$.  The difference
between $\mw^{exp}$ and $\mw^{fit}$ is slightly more than one standard
deviation. In view of possible future improvements in the experimental
accuracy at the  LHC, it is therefore worthwhile to reconsider the indirect
determination of $\mw$ and in particular its theoretical uncertainty.
In this paper we  study the $\mw-\mz$ interdependence at the
two-loop level following a path different from the one employed so
far, i.e.\ the two-loop determination of $\dr$ in the OS scheme, and we
 critically re-examine the overall theoretical uncertainty of the
SM prediction of $\mw$.

The \bms\ formulation of the radiative corrections in the SM was
developed in Refs.~\cite{Si89,FS, DFS} and provides an alternative way
to address the $\mw-\mz$ interdependence.  In this framework the gauge
coupling constants are defined as $\bms$ quantities, while all the
masses are interpreted as pole quantities\footnote{We generically refer
  to this approach as \bms\ scheme, although it is actually a hybrid
  OS-\bms\ scheme.}.  All gauge couplings are then reexpressed in terms of the \bms\ Weinberg angle $\hat{\theta}_{W} (\mu)$ and the
\bms\ electromagnetic coupling $\alc (\mu)$, defined at the 't-Hooft
mass scale $\mu$, usually chosen to be equal to $\mz$. The important
feature of these two \bms\ parameters is that they are constructed to
include all reducible contributions, i.e.\ the iteration of lowest
order terms. In particular, $\alc (\mz)$ automatically incorporates
the ${\cal O}(\alpha^n \ln^n\mz/m_f)$ contributions, while $\sincur
\equiv \scs$ is free  of 
the ${\cal O}((\alpha \mt^2/\mw^2)^n)$
contributions that in the OS scheme are induced by the renormalization
of the OS $\theta_W$ angle. Similarly, the on-shell masses of the vector bosons
automatically absorb the non-decoupling contributions of heavy particles to 
their self-energies. It follows that in this hybrid scheme
higher order effects  are expected to be  better under control with respect 
to the OS  or a pure \bms\ schemes.

In the \bms\ formulation the $\mw-\mz$ interdependence is expressed in
terms of three parameters $\drcarw$, $\Delta \alc$ and $\rhoh$,
 defined by 
\bea&& \frac{G_\mu}{\sqrt{2}} = \frac{\pi \alc
  (\mz)}{2 \mw^2 \scs} \left[ 1 + \drcarw \right], \label{drcarw}
\qquad \alc (\mz)= \frac{\alpha}{1-\Delta \alc (\mz)}, \nonumber\\ &&
\qquad \qquad \qquad \qquad\rhoh = \frac{\mw^2}{\mz^2 \ccs} =
\frac{c^2}{\ccs} \, 
\label{rhoc}
\eea
where $\ccs = 1- \scs$. Eqs.(\ref{rhoc}) allow for an
iterative evaluation of $\scs$ from $\mz, \alpha, G_\mu$:
\be
        \scs = \frac{1}{2} \left\{
         1 - \left[ 1 - \frac{4 \hat{A}^2}{\mz^2 \rhoh} (1+\drcarw) \right]^{1/2}
                            \right\} ~,
\label{sincur}
\ee
where $\hat{A}= (\pi \alc (\mz)/(\sqrt{2} G_\mu))^{1/2}$. The analogue
 for $\mw$ reads
\be
        \mw^2 = \frac{\rhoh\, \mz^2}{2} \left\{
         1 + \left[ 1 - \frac{4 \hat{A}^2}{\mz^2 \rhoh} (1+\drcarw) \right]^{1/2}
                            \right\}~ .
\label{mwcur}
\ee

The present knowledge of $\alc(\mz),\, \drcarw,\, \rhoh$ can be
summarized as follows: a complete EW two-loop calculation for
$\alc(\mz)$ was presented in Ref.\cite{DV}. The other two-parameters
are not known at the same level of accuracy: they are only known at
the second order in the heavy top expansion,  i.e.\ up to the
two-loop ${\cal O}(\alpha^2 \mt^2/\mw^2)$ contributions \cite{DGV,
  DGS}. In this paper we  upgrade the \bms\ calculation at the
full  two-loop level presenting the complete ${\cal O}(\alpha
\as)$ and ${\cal O}(\alpha^2 )$ determination of $\alc(\mz),\,
\drcarw,\, \rhoh$ augmented by the known three-loop corrections.

The precise knowledge of $\mw$ in the \bms\ framework allows us to
estimate the uncertainty of the $\mw$ prediction in two different
ways: i) from the scale dependence of our \bms\ result by varying the 't 
Hooft mass scale in a large  interval between 50 and 500 GeV. ii) 
From the scheme dependence by comparing our result in the \bms\ scheme with 
the known result in the OS scheme present in the literature.
 
As a byproduct of our \bms\  calculation, we  also obtain the
values of the \bms\ gauge couplings at the weak scale with a two-loop
precision. The latter can be used as initial conditions for
studies of the renormalization group evolution.

The paper is organized as follows: in the next section we outline 
our computation. Section 3 discusses
the two-loop determination of  $\alc(\mz),\, \drcarw,\, \rhoh$. Section
4 contains our results for $\alc (\mu), \, \sincur(\mu)$ and  $\mw$.
In the last section  we discuss  the uncertainty
on the theoretical determination of $\mw$  and present our conclusions. 

\section{Outline of the computation}
\label{outline}
In this section we first extend at the two-loop level
the \bms\ framework developed at one-loop in Refs.\cite{Si89,FS, DFS}.
Then some technical details concerning our computation are outlined.

The parameters that in our computation require a two-loop
renormalization
are  the two gauge couplings, $g,\: g^\prime$,  
and the masses of the gauge bosons. Actually, as the
gauge sector of the SM is described by only 3 parameters,
$g,\: g^\prime$ and $v$, the vacuum expectation value (vev) of the Higgs field,
once the two gauge couplings are  defined as \bms-subtracted quantities,
one needs to define the mass of only one gauge boson, either the 
$W$ or the $Z$, while the renormalized mass of the other boson is obtained using
the bare relation $\mzo = \mwo/\cos\!\theta_{{W}_0}$. We first identify our
vev  as the minimum of the  radiatively corrected scalar potential. The 
latter  implies that all tadpole  contributions are cancelled by  a
tadpole counterterm and that tadpole  diagrams do not enter in our computation. 
We choose to  define our renormalized $W$ mass, $\mw$, as a pole
quantity fixing our third renormalization condition. 
Our renormalized $Z$ mass,  $\hat{m}_{\scriptscriptstyle Z}$, is a derived quantity
identified with $\hat{m}_{\scriptscriptstyle Z} \equiv \mw/\hat{c}$. The
use of the experimental quantity $m_{\scriptscriptstyle Z}^{exp}$ as 
input  in eqs.\,(\ref{sincur}, \ref{mwcur}) requires the derivation, at the two-loop level, of the relation between 
$\hat{m}_{\scriptscriptstyle Z}$ and $m_{\scriptscriptstyle Z}^{exp}$.

According to our choice of pole mass for the $W$ boson, 
at the one-loop level $\mw$ can be directly 
identified with  $m_{\scriptscriptstyle W}^{exp}$ and its the counterterm, 
$\delta \mw^2$,  is  given by:
\be
\delta^{(1)} \mw^2 = {\rm Re} \,\aww^{(1)} (\mw^2)
\label{Wcontro}
\ee
where, in general, $A_{XY}(q^2)$ is the term proportional to $g^{\mu \nu}$ in 
the $XY$ 
self-energy and  the superscript  indicates the loop order. 
Because of our condition on 
the cancellation  of the tadpoles, no 
tadpole term is  included in eq.\,(\ref{Wcontro}) 

At the two-loop level  the definition of a pole  mass for an unstable gauge 
boson  presents some subtlety in its relation with the corresponding experimental quantity. Since the beginning of the 
nineties it was noticed \cite{Si91} that,  beyond  one-loop order,  
there is a difference between the mass  defined as the pole of the real 
part of the propagator (labelled $m$), or as the real part of the 
complex  pole of the S matrix, $M$ in the following. We recall here the 
discussion on the $Z$ 
mass developed in Ref.\,\cite{Si91} that can also be applied to the $W$ case. 
The former definition leads to the $Z$ mass counterterm 
\be
\delta \mz^2 = {\rm Re}\, \azz (\mz^2)
\label{Zcontro1}
\ee
and gives rise to a renormalized mass that, at the two-loop level,
is gauge-dependent if the r.h.s.\ of eq.\,(\ref{Zcontro1}) is 
evaluated in a gauge where the  gauge parameter satisfies
\mbox{$\xi < (4 \cos^2 \theta_W)^{-1}$},
while it is still gauge-independent if the evaluation is performed with
\mbox{$\xi \geq (4 \cos^2 \theta_W)^{-1}$}.
Let us now denote by  $\overline{s}$ the position of the complex pole of the
$Z$ propagator. Hence
\be
\overline{s} =\Mzo^2 + \azz(\overline{s}),
\label{polepos}
\ee
where $\Mzo$ is the bare mass. The complex pole  definition of 
the renormalized mass and width of the 
$Z$ boson follows immediately,
\be
\overline{s} =\Mz - i \Mz \Gz,
\label{compmass}
\ee
and gives rise to a two-loop mass counterterm given by
\be
\delta^{(2)} \Mz^2 = {\rm Re} \,\azz^{(2)} (\Mz^2) + {\rm Im}\, \azz^\prime (\Mz^2)
\Mz \Gz~.
\label{Wcontro2p}
\ee

The $Z$  boson mass defined according to 
the real part of the complex pole of the $S$ matrix
generates a fixed-width Breit-Wigner behavior of the total cross section 
while $m_{\scriptscriptstyle Z}^{exp}$ is extracted  using a  Breit-Wigner
parametrization with an energy dependent width. This  introduces
a mismatch among
the parameters entering the r.h.s.\ of eq.\,(\ref{compmass}) and their 
experimental counterparts that is corrected by \cite{BLRS, Si91}:
\be
\Mz =m_{\scriptscriptstyle Z}^{exp} \left[ 1+ \left(  
\frac{\Gz^{{exp}}}{m_{\scriptscriptstyle Z}^{{exp}}} \right)^2 \right]^{-1/2},~~~~~~~~~
\Gz = \Gz^{exp} \left[ 1+  \left(
\frac{\Gz^{{exp}}}{m_{\scriptscriptstyle Z}^{{exp}}}\right)^2 \right]^{-1/2}~.
\label{relMG}
\ee
On the other hand, $\mz$ defined as the pole of the real part of the propagator
can be directly identified with $m_{\scriptscriptstyle Z}^{exp}$ if one works at 
the two-loop level evaluating eq.\,(\ref{Zcontro1}) in a gauge
with  \mbox{$\xi \geq (4 \cos^2 \theta_W)^{-1}$}.

We decided to identify our renormalized $W$ mass directly with the quantity extracted
experimentally. According to the above discussion
this fixes $ \delta^{(2)} \mw^2$ to be
\be
\delta^{(2)} \mw^2 = {\rm Re}\, \aww^{(2)} (\mw^2)
\label{Wcontro2}
\ee
with the understanding that the r.h.s.\ of eq.\,(\ref{Wcontro2}) has to be
evaluated in a gauge where spurious gauge-dependent terms do not arise.
The same condition  applies to the relation between  
$\hat{m}_{\scriptscriptstyle Z}$ and $\mz$, which is identified with
$m_{\scriptscriptstyle Z}^{exp}$. 
We fulfill it by evaluating ${\rm Re} \aww^{(2)} (\mw^2)$ and
${\rm Re} \azz^{(2)} (\mz^2)$ in the $\xi=1$ Feynman gauge.
We stress that with our choice the $\mw$ prediction of eq.\,(\ref{mwcur})
can be {\it directly} compared with 
$  m_{\scriptscriptstyle W}^{{exp}}$. 
The other possible definition of the $W$ mass, 
$\Mw$, requires instead the correction factor of eq.\,(\ref{relMG})
before it can be compared with  
$  m_{\scriptscriptstyle W}^{exp}$.

The other mass parameters that enter our computation  require only a one-loop
definition. We define the Higgs, top and bottom masses as pole
quantities. The bottom mass is set different from zero only in the
one-loop contribution and in the ${\cal O}(\alpha \as)$ corrections.
All other quarks are taken massless. The leptons are also taken massless
except for  the evaluation of $\alc$ where the experimental values in the
Particle Data Group \cite{PDG} have been used.

We conclude this section outlining  some technical details concerning our 
computation.
All the diagrams entering the calculation of $\alc(\mz),\, \drcarw,\, \rhoh$
were generated using the Mathematica package {\sc Feynarts}~\cite{Feynart}.
The reduction of the two-loop diagrams to scalar integrals was done using 
the code {\sc   Tarcer} \cite{Mertig:1998vk}  which uses the 
 algorithm by Tarasov \cite{Tarasov} and is now part of the {\sc
 Feyncalc}~\cite{Feyncalc} package. In order to extract the 
vertex  and box contributions in $\drcarw$ from the relevant
diagrams, we used the  projector presented in Ref.\,\cite{ACOV}.
After the reduction to scalar integrals we were left with
the evaluation of two-loop vacuum integrals and
two-loop self-energy diagrams at external momenta different from zero.
The former integrals were evaluated analytically using the results
of Ref.\,\cite{DT}. The latter ones  were instead
reduced to the set of loop-integral basis functions introduced
in Ref.\,\cite{Martinint}. The evaluation of the basis functions was 
done numerically using the code {\sc TSIL}~\cite{MartinTSIL} that, according
to the authors, reaches a relative accuracy better than $10^{-10}$ in the
evaluation of integrals without large hierarchies in the masses.

All our results  were obtained  in the  $R_\xi$ gauge with $\xi=1$ and
cross-checked in the $\xi=1$ background field method (BFM) gauge. 
The two-point function of a particle, i.e.\ the sum of the self-energy and of 
the tadpole diagrams, when evaluated on-shell represent a physical amplitude 
and 
must be gauge-invariant. Enforcing the cancellation of the tadpoles, we
verified that the sum of the one-particle-irreducible and counterterms 
diagrams in ${\rm Re} \aww^{(2)} (\mw^2)$ and ${\rm Re} \azz^{(2)} (\mz^2)$
gives  the same result in the two gauges.

\section{Two-loop determination of  $\alc(\mz),\, \drcarw,\, \rhoh$}
In this section we present  the two-loop contributions to the three radiative 
parameters of the \bms\ scheme. To properly identify the two-loop 
contribution to these parameters  the exact specification 
of the corresponding one-loop result is needed.
In the Appendix we report the one-loop expressions for 
 $\alc(\mz),\, \drcarw,\, \rhoh$ that we employed in our computation. 

\subsection{$\alc(\mz)$}
The evaluation of the
electromagnetic coupling in  the $\bms$ scheme at the two-loop level was
discussed in Ref.\,\cite{DV}. Here we just recall the main features of that
analysis and update the QCD corrections.

\begin{table}[t]
\begin{center}
$$\begin{array}{rcl}
\mz &=& 91.1876\pm0.0021\, \gev \\
\mh &=& 125.15 \pm 0.24\, \gev \\
\mt &=&173.34\pm 0.76_{exp}\, \pm 0.3_{th}\,\gev  \\
m_e &=& 0.510998928\pm 0.000000011\,\mev \\
m_{\mu} &=& 105.6583715\pm 0.0000035\,\mev \\
m_{\tau} &=& 1776.82  \pm 0.16\,\mev \\
m_b &=& 4.8\pm 0.3 \,\gev \\
G_\mu &=& 1.1663781\pm 0.0000006 \times 10^{-5}\gev^{-2} \\
\as(\mz) &=& 0.1184\pm 0.0007\\
\Dah &=& 0.02750 \pm 0.00033 
\end{array}$$
\caption{\label{table1} Experimental input  values used in our analysis}
\end{center}
\end{table}%

The analysis starts from the observation that in the Feynman BFM gauge
the renormalization of the electric charge is given only by self-energy
diagrams making manifest the possibility of a Dyson summation. From the
relation between the bare and the renormalized electric charge defined
at zero momentum transfer
\be
    e^2 = \frac{e^2_0}{ 1 - e^2_0\,\pigg(0) }~,
\label{eq1SM}
\ee
where $\Pi_{\gamma \gamma}$ is related to the transverse part of
the photon self-energy $\agg (q^2) $ by
\be
\agg (q^2) = q^2 \,e^2_0\,\pigg (q^2)~
\ee
it is easy to derive the relation between $\alpha =(137.035999074)^{-1}$ 
and the electromagnetic coupling in  the $\bms$ scheme at the scale $\mu$
\be
\alc (\mu) = \frac{\alpha}{ 1 -  \Delta \alc (\mu)}
\label {alfacur}
\ee
with 
\be
\Delta \alc (\mu) = -4 \,\pi\, \alpha\, \pigg(0)|_{\ms}
\label {dalpha}
\ee
where \bms\ is denoting the \bms\ renormalization. As we are interested in 
the evaluation of $\alc(\mu)$ in the SM at a scale  below 
$\mu=\mt$ we do not apply the decoupling of the top contribution 
from 
$\pigg(0)$. 
  
The vacuum polarization function in eq.\,(\ref{dalpha}) can be organized into 
the sum of a bosonic 
and a fermionic contribution, the latter defined as arising from diagrams where the 
external photons couple both to fermions, 
\be
\pigg(0) = \piggp{f}(0) + \piggp{b}(0)~.
\label{pis}
\ee
The fermionic contribution can be further split  into a leptonic part,
$\piggp{l}$, a perturbative quark contribution, $\piggp{p}$, and a 
non-perturbative one, $\piggp{5}(0)$. The latter, associated to
diagrams in which a light quark couples to the external photons with no heavy 
masses circulating  in the loops,  can be 
related to  the hadronic contribution to the vacuum polarization
$\Dah \equiv  4 \pi \alpha 
\left({\rm Re}\,\piggp{5}(\mz^2) -\piggp{5}(0)\right)$ so that 
\bea
\piggp{f}(0) &=& \piggp{l}(0)  +  \piggp{p}(0) + \piggp{5}(0) \nonumber\\
             &=& \piggp{l}(0)  +  \piggp{p}(0) + 
              \left(\piggp{5}(0) - {\rm Re}\, \piggp{5}(\mz^2)\right) +
               {\rm Re}\,\piggp{5}(\mz^2)~.
\label{pif}
\eea
The hadronic contribution can be obtained from the experimental data on the 
cross section in $e^+ e^- \rightarrow hadrons$ by using a dispersion relation.
Two recent  evaluations of $\Dah$ report very consistent results: 
$\Dah = (275.7 \pm 1.0) \times 10^{-4}$
\cite{DHMZ},  $\Dah = (275.0 \pm 3.3) \times 10^{-4}$ \cite{BP}. 
We use the latter  as reference value in our calculation.
The  $\piggp{p}$ term in eq.\,(\ref{pif}) includes the top contribution to 
the vacuum polarization plus the 
two-loop   diagrams in which a light quark  couples internally
to the $W$ and $Z$ bosons. This contribution, as well as 
${\rm Re}\,\piggp{5}(\mz^2)$, can be safely analyzed perturbatively.

The one-loop contribution to 
$\Delta \alc^{p} (\mz) \equiv \Delta \alc (\mz)- \Dah$ is reported in 
eq.\,(\ref{Dalpha1})
of the Appendix. The higher order contributions to
$\Delta \alc^{p} (\mz)$ are presented here as a simple formula that
parametrizes the full result in terms of the top and  the Higgs masses, 
the strong coupling, and $\scs$:
\be
\Delta \alc^{p,\,h.o.} (\mz) =10^{-4}\left( b_0 + 
        b_1 ds + b_2 dT + b_3 dH  + b_4 da_s \right)
\label{alphapar}
\ee
where
\bea
&&ds = \!\left(\frac{\scs}{0.231}-1\right), \qquad
dT = \ln\left(\frac{\mt}{173.34\,{\rm GeV}}\right),\, \nonumber\\
&&dH = \ln\left(\frac{\mh}{125.15\,{\rm GeV}}\right),\, \qquad
da_s = \left(\frac{\as(\mz)}{0.1184}-1\right)
\eea
with
\be
b_0=1.751181 ~~~ b_1=-0.523813, ~~~ b_2=-0.662710,~~~ 
b_3=-0.000962,~~~b_4= 0.252884~.
\label{alphacoeff}
\ee
Eq.\,(\ref{alphapar}) includes the ${\cal O} (\alpha)$ contribution
to $\piggp{b}(0) +\piggp{l}(0)+\piggp{p}(0)$ plus  the  ${\cal O} (\as)$
corrections to $\piggp{p}(0)$ and the ${\cal O} (\as,\, \as^2)$
corrections to ${\rm Re}\,\piggp{5}(\mz^2)$ \cite{QCD3la}. It 
approximates the exact result
to better than $0.045 \%$ for $\scs$ in the interval $(0.23-0.232)$ when
the other parameters  in  Eq.\,(\ref{alphapar}) are varied simultaneously
within a $3\sigma$ interval around their central values, given in  Table \ref{table1}.

\subsection{\drcarw}
The radiative parameter $\drcarw$ enters  the relation between the Fermi
constant and the $W$ mass.  We recall that the Fermi constant 
is defined in terms of the muon lifetime $\tau_\mu$ as computed
in an effective  4-fermion $V-A$ Fermi theory supplemented by QED interactions: 
\be 
\label{eq:taumu}
\frac{1}{\tau_\mu} = \frac{G_\mu^2 m_\mu^5}{192\pi^3} F\left(\frac{m_e^2}{m_\mu^2}\right) 
(1 + \Delta q)\left(1+\frac{3m_\mu^2}{5 \mw^2}\right) \ ,
\ee
where $F(\rho)=1-8\rho+8\rho^3-\rho^4-12\rho^2\ln\rho=0.9981295$ 
(for $\rho=m_e^2/m_\mu^2$) is the phase space factor and
$\Delta q=  \Delta q^{(1)} + \Delta q^{(2)}=(-4.234+0.036)\times10^{-3}$
are the QED corrections computed at one~\cite{Kinoshita:1958ru} and two 
loops~\cite{Deltaq}. The calculation
of $ \drcarw$ requires  the subtraction of the QED corrections,
 matching the result in the SM with that in the Fermi theory which is 
renormalizable to all orders in the
electromagnetic interaction but to lowest order in $G_\mu$. 
In the limit of vanishing fermion masses the matching requires
just the calculation in the SM with the contribution of the Fermi effective 
theory to the Wilson coefficient 
vanishing.

The muon-decay amplitude at the two-loop level can be written as
\be
\frac{G_\mu}{\sqrt{2}} = \frac{g_0^2}{8 \mwo^2} \left\{1 - 
\frac{\aww}{\mwo^2} + V_W + \mwo^2 B_W 
 + \left(\frac{\aww}{\mw^2}\right)^2 - \frac{\aww V_W}{\mw^2}
\right\}
\label{Gmu}
\ee 
where $g_0$ is the unrenormalized $SU(2)$ coupling, $\mwo$ is the
unrenormalized $W$ mass, $\aww \equiv \aww (0)$, and $V_W$ and $B_W$ are the 
relevant vertex and box contributions to $\mu$-decay. 
Performing the shift
$\mwo^2 \to \mw^2 - \delta \mw^2$, and working at the two-loop order
we arrive at
\bea
 \frac{G_\mu}{\sqrt{2}}  &=& \frac{g_0^2}{8 \mw^2} \left[ 
1 + \frac{\delta^{(1)} \mw^2}{\mw^2} -\frac{\aww^{(1)}}{\mw^2} + E^{(1)} +
\frac{\delta^{(2)} \mw^2}{\mw^2} -\frac{\aww^{(2)}}{\mw^2} + E^{(2)} 
 \right. \nonumber \\
&& \left. + \aww^{(1)} B^{(1)}_W + \left( \frac{ \delta^{(1)} \mw^2}{\mw^2}-
  \frac{\aww^{(1)}}{\mw^2}\right)\left( \frac{ \delta^{(1)} \mw^2}{\mw^2}-
   \frac{\aww^{(1)}}{\mw^2} +E^{(1)}\right) \right]
\label{Gmu2}
\eea
where the superscript indicated the loop order and
$ E^{(i)} \equiv V^{(i)}_W + \mwo^2 B^{(i)}_W$.  Performing an $\bms$ 
renormalization of the $SU(2)$ and $U(1)$ couplings we write
\be
 \frac{G_\mu}{\sqrt{2}}  = \frac{\pi \alc (\mz)}{2 \mw^2 \scs} \left[ 1 +
\drcarw \right]
\label{Drcarw}
\ee
with
\bea
\drcarw & = & \drcarwu + \drcarwd ~,\\
\drcarwu &=& \left. \frac{{\rm Re} \aww^{(1)} (\mw^2)}{\mw^2} -
            \frac{\aww^{(1)}}{\mw^2} + E^{(1)} \right|_{\ms}~,
\label{drcarw1}\\
\drcarwd &=&
\frac{{\rm Re} \aww^{(2)}( \mw^2)}{\mw^2} -\frac{\aww^{(2)}}{\mw^2} + E^{(2)} 
  +\delta^{\epsilon} \drcarwu  + \aww^{(1)} B^{(1)}_W \nonumber \\
&& + \left.
\left( \frac{ {\rm Re} \aww^{(1)}( \mw^2)}{\mw^2}-
  \frac{\aww^{(1)}}{\mw^2}\right)
\left( \frac{ {\rm Re} \aww^{(1)}( \mw^2)}{\mw^2}-
   \frac{\aww^{(1)}}{\mw^2} +E^{(1)}\right)  \right|_{\ms}~.
\label{drcarw2}
\eea
where \bms\ in this case denotes both the \bms\ renormalization 
and the  choice $\mu = \mz$ for the 't Hooft mass scale; 
$\delta^{\epsilon} \drcarwu$ is the finite contribution related to the
$\epsilon =(4-d)/2$ part of $\drcarwu$, $d$ being the dimension of the 
space-time.  

We note that
the definition of $\drcarw$ in eq.\,(\ref{Drcarw}) differs from the original
proposal in Ref.\,\cite{FS},
\be
 \frac{G_\mu}{\sqrt{2}}  = \frac{\pi \alpha }{2 \mw^2 \scs} \frac1{ 1 -
\drcarw}  \label{drcarwo}~. 
\ee
In eq.\,(\ref{drcarwo}) the relation between $G_\mu$ and
$\mw$ is expressed in terms of $\alpha$ and  the mass singularity
corrections  are directly included in $\drcarw$ and their resummation is 
achieved via the replacement
\be
1+ \drcarw \to \frac1{1-\drcarw}~.
\label{replace}
\ee
This is clearly different from eq.\,(\ref{Drcarw}) where they are absorbed in
$\alc(\mz)$.
The replacement
(\ref{replace}) used in eq.\,(\ref{drcarwo}) introduces spurious two-loop
and higher-order contributions, numerically quite small.  The 
use of   eq.(\ref{Drcarw}) allows us instead to control directly the resummation of the
various contributions. 

An explicit expression for $\drcarwu$ is reported in eq.(\ref{Drcur1})
of the Appendix where the distinction between $\ccs$ and $\cq$ is kept.
The higher order contributions to $\drcarw$ are presented again in  a simple
formula that approximates the exact result
to better than $0.035 \%$ for $\scs$ on the interval $(0.23-0.232)$ when the 
other parameters   are varied simultaneously
within a $3\sigma$ interval around their central values. We find  
\be
\drcarw^{\,h.o.} (\mz) =10^{-4}\left( r_0 + r_1 ds + r_2 \,dT + r_3 \,dH +
r_4\, da_s \right)
\label{drcurpar}
\ee
with
\be
r_0=-2.8472779, ~~~ r_1=1.620742, ~~~ r_2=1.773226,~~~ 
r_3=-0.364310,~~~r_4= 1.137797~.
\label{drcurcoeff}
\ee
Eq.(\ref{drcurpar}) includes, besides the $\drcarwd$ contribution from 
Eq.(\ref{drcarw2}), the complete ${\cal O} (\alpha \as)$ corrections and
the first two subleading terms in the heavy top expansion of the three-loop 
${\cal O} (\alpha \as^2 )$ corrections

\subsection{\rhoh}
The  relation between the Weinberg angle in the \bms\ formulation and its 
OS counterpart is encoded in the parameter $\rhoh$ defined as
\be
        \rhoh = \frac{c^2}{\ccs} = \frac{\mw^2}{\mz^2 \ccs} 
\ee
whose tree-level value is equal to 1. From the relation
\be
\frac{\mwo}{\mzo} \equiv c_0^2 = c^2 -c^2  \frac{\delta \mw^2}{\mw^2} +c_0^2 
  \frac{\delta \mz^2}{\mz^2} = \ccs - \delta \ccs
\label{rhocd}
\ee
with $ \delta \mz$ given by eq.\,(\ref{Zcontro1})   and $ \delta \ccs$
 the counterterms for $\ccs$,  it is easy to derive
\be
        \rhoh = \frac{1}{( 1 - Y_{\ms} )} \, .
\label{rhocd1}
\ee
with
\be
Y =\frac{\delta \mw^2}{\mw^2} - c_0^2   \frac{\delta \mz^2}{\mw^2}~.
\label{Yms}
\ee
In eq.\,(\ref{rhocd1}) \bms\ denotes both the \bms\ renormalization 
and the  choice $\mu = \mz$ for the 't Hooft mass scale. Indeed  
the structure of the $1/\epsilon$ poles in $ \delta \ccs$
is identical to that of the combination of the $W$ and $Z$ mass counterterms
in eq.\,(\ref{rhocd}) once  the $1/\epsilon$ poles in  ${\delta^{(1)} \mw^2}$
and ${\delta^{(1)} \mz^2}$ are expressed in terms of 
\bms\ quantities.

The two-loop counterterm $\delta^{(2)} \mz^2$ includes also the contribution
from the mixed $\gamma\, Z$ self-energy
or 
\be
\delta^{(2)} \mz^2  =  {\rm Re} \left[ \azz^{(1)} ( \mz^2) + \azz^{(2)}( \mz^2) +
                    \left( \frac{\agz^{(1)} (\mz^2)}{\mz^2} \right)^2\right]
\label{dmz} 
\ee
so that $Y_{\ms}$ up to the two-loop level reads
\bea
        Y_{\ms} & = &  Y^{(1)}_{\ms} + Y^{(2)}_{\ms} \, ,\\
        Y^{(1)}_{\ms}  & = &  {\rm Re} \left[
         \frac{\aww^{(1)}(\mw^2)}{\mw^2} - \ccs \frac{\azz^{(1)}(\mz^2)}{\mw^2} 
                \right]_{\ms} \, , \\
 Y^{(2)}_{\ms} & = &  {\rm Re} \left[
         \frac{\aww^{(2)}(\mw^2)}{\mw^2} -  \frac{\azz^{(2)}(\mz^2)}{\mz^2} + 
             \left(  \frac{\agz^{(1)}}{\mz^2} \right)^2
                \right]_{\ms}~.
\label{Yms2l}
\eea

The one-loop contribution to 
$Y_{\ms}$ is reported in eq.\,(\ref{Yms1l})
of the Appendix. As before we give the higher order terms via a simple formula: 
\be
Y_{\ms}^{\,h.o.} (\mz) =10^{-4}\left( y_0 + 
       y_1 ds + y_2 dt + y_3 dH  + y_4 da_s \right)
\label{ymspar}
\ee
where
$dt = [(\mt/173.34 \, {\rm GeV})^2-1]$ and
\be
y_0=-18.616753 ~~~ y_1=15.972019, ~~~ y_2=-16.216781,~~~
y_3=  0.0152367,~~~y_4= -13.633472~.
\label{ymscoeff}
\ee
Eq.\,(\ref{ymspar}) includes, besides the $Y^{(2)}_{\ms}$ contribution from 
eq.\,(\ref{Yms2l}), the complete ${\cal O} (\alc \as)$ corrections,
the leading three-loop ${\cal O}(\alc \as^2 \mt^2)$ contribution 
\cite{QCD3lrho} and the subleading ${\cal O}(\alc^3 \mt^6)$ and
${\cal O}(\alc^2 \as\mt^4)$ \cite{FKSV}. It approximates the exact result
to better than $0.075 \%$ for $\scs$ on the interval $(0.23-0.232)$ when 
the other parameters  in  eq.\,(\ref{ymspar}) are varied simultaneously
within a $3\sigma$ interval around their central values.

\section{Results}
In this section we report our results for $\alc, \: \sincur$ and $\mw$.
All results are presented as simple parameterizations in terms of the relevant 
quantities whose stated validity refers to a simultaneous  variation of the
various parameters  within a $3 \sigma$ interval around their central values 
given in Table \ref{table1}.  As a general strategy for the evaluation of the
two-loop contributions, where $\ccs$ can be identified with $\cq$, we have
replaced in all the two-loop terms $\mw$ with $\mz \hat{c}$. This choice 
gives rise to the weakest $\mu$-dependence in $\mw$. 

\begin{table}[t]
\begin{center}
\begin{tabular}{c|c|c|}
    & $\mu=\mz$   & $\mu = \mt$  \\
\hline
$a_0$ & $ (128.13385)^{-1}$ & $(127.73289)^{-1}$ \\ 
\hline
$a_1$ &-0.00005246 & -0.00005267 \\ 
\hline
$a_2$ &-0.01688835 & 0.02087428 \\
\hline 
$a_3$ & 0.00014109 & 0.00168550 \\ 
\hline
$a_4$ &0.22909789 & 0.23057967 \\ 
\hline
\end{tabular}\hspace{2cm}
\begin{tabular}{c|c|c|}
   & $\mu=\mz$ & $\mu= \mt$ \\
\hline
 $s_0$ & 0.2314483 & 0.2346176\\ 
\hline
 $s_1$ & 0.0005001 & 0.0005016\\
\hline
 $s_2$ & -0.0026004 & -0.0001361\\
\hline
 $s_3$ & 0.0000279 & 0.0000514\\
\hline
 $s_4$ & 0.0005015 & 0.0004686\\
\hline
 $s_5$ & 0.0097431 & 0.0098710\\
\hline
\end{tabular}
\caption{\label{table2} Coefficients for the parameterization of
$\alc (\mu) $ (left table, eq.\,(\ref{par1}) in the text) and $\sincur (\mu)$ 
(right table, eq.\,(\ref{par2}) in the text). } 
\end{center}
\end{table}

The two-loop computation of the \bms\ electromagnetic coupling from 
eq.\,(\ref{alfacur}) and of $\sincur$ from eq.\,(\ref{sincur}) can be 
summarized by the following  parameterizations
\bea
\alc (\mu)& = &  a_0 + 10^{-3} \left( a_1 dH + a_2 dT + a_3 da_s +a_4 d a^{(5)} 
        \right) \label{par1}\\
\sincur(\mu) & = & s_0 + s_1 dH + s_2 dt + s_3 dHdt  + s_4 da_s +
s_5 da^{(5)}
\label{par2}
\eea
where $da^{(5)}=[\Dah/0.02750-1]$ and the $a_i$ and $s_i$ coefficients are
reported in Table \ref{table2} for two different values of the
 scale $\mu$. Eq.(\ref{par1}) 
approximates the exact result to better than $1.1\times 10^{-7}$ ($1.2 \times 10^{-7}$) for $\mu = \mz$ ($\mu= \mt$),
while eq.\,(\ref{par2})  approximates the exact 
result to better than $5.1 \times 10^{-6}$ ($6.2 \times 10^{-6} $) for $\mu = \mz$ ($\mu= \mt$). 

From our results on $\alc$ and $\scs$ it is easy to obtain the values
of the $g$ and $g'$ coupling constants at the weak scale, usually
identified with $\mt$. They can be  taken as starting points
in the study of the evolution of the gauge couplings via
Renormalization Group Equations (RGE) in Grand Unified Models and in the 
analysis of the stability of the Higgs potential in the SM. 
Ref.\,\cite{Buttazzo} reports the values of the gauge coupling 
constants at the $\mu=\mt$ scale,   $g(\mt) =0.64822$ and
$g'(\mt) = 0.35760$, obtained using a complete  calculation of the two-loop 
threshold corrections in the SM.   Here we find 
$g(\mt) = 0.647550 \pm 0.000050$ and $g' (\mt) =0.358521\pm 0.000091$. 
The difference between the two results, which  should be a three-loop effect,  is  more
sizable than expected.
However, 
the results of Ref.\,\cite{Buttazzo} were obtained using as input
parameters $G_\mu$ and the experimental values of $\mz$ and $\mw$,  while
our result is obtained with a different set of input parameters, i.e.\
$G_\mu, \, \alpha$ and $\mz$. In our calculation $\mw$ is a derived quantity calculable from
eq.\,(\ref{mwcur}). Moreover, as shown below, our prediction for $\mw$
is not in perfect agreement with the  present experimental determination and
therefore the gauge couplings extracted using  the two different sets
of inputs parameters show some discrepancy. Indeed, using our prediction for $\mw $   in the results of Ref.\,\cite{Buttazzo}  instead
of the experimental result,  
we find that the difference between the $g$ ($g'$) computed in the two methods is one order of magnitude smaller than the two-loops correction and two orders smaller than the one-loop correction to $g$ ($g'$).

\begin{table}[t]
\begin{center}
\begin{tabular}{c|c|c|}
    & $124.42 \le \mh \le 125.87$ GeV   & $50 \le \mh \le 450$ GeV  \\
\hline
$w_0$ & 80.35712  & 80.35714 \\ 
\hline
$w_1$ & -0.06017  & -0.06094 \\ 
\hline
$w_2$ &  0.0      & -0.00971 \\
\hline 
$w_3$ &  0.0      & 0.00028 \\
\hline 
$w_4$ &  0.52749  &  0.52655\\ 
\hline
$w_5$ & -0.00613  & -0.00646\\ 
\hline
$w_6$ & -0.08178  & -0.08199 \\ 
\hline
$w_7$ & -0.50530  & -0.50259 \\ 
\hline
\end{tabular}
\caption{\label{table3} Coefficients of the $\mw $ parameterization in
 eq.\,(\ref{parw}). The left column contains the
coefficients that cover a  variation of $\mh$ around its central value,
while the right one applies to the case $50 \le \mh \le 450$ GeV . } 
\end{center}
\end{table}

The two-loop determination of the $W$ mass in the \bms\ framework 
from eq.\,(\ref{mwcur}) can be parameterized as follows
\be
\mw^2 = w_0 + w_1 dH+ w_2 dH^2 +w_3 dh + w_4 dt + w_5 dHdt  + w_6 da_s +
             w_7 da^{(5)}
\label{parw}
\ee
with $dh = [(\mh/125.15 \, \gev)^2 -1]$.
The $w_i$ coefficients are reported   in Table \ref{table3} for  $\mu=\mz$.   
Two different cases are considered.
In the left column the coefficients  refer to the standard case of a 
simultaneous variation of all parameters within a $3\sigma$ interval around 
their central values. The right column applies to the case where all
parameters but the Higgs mass are varied within a $3\sigma$ interval
while the latter is varied between
50 and 450 GeV. In the two cases the formula (\ref{parw}) approximates the exact result to better than  $0.11$\,MeV and $0.5$\,MeV, respectively.


\begin{figure}
\begin{center}
\includegraphics[width=0.83\textwidth]{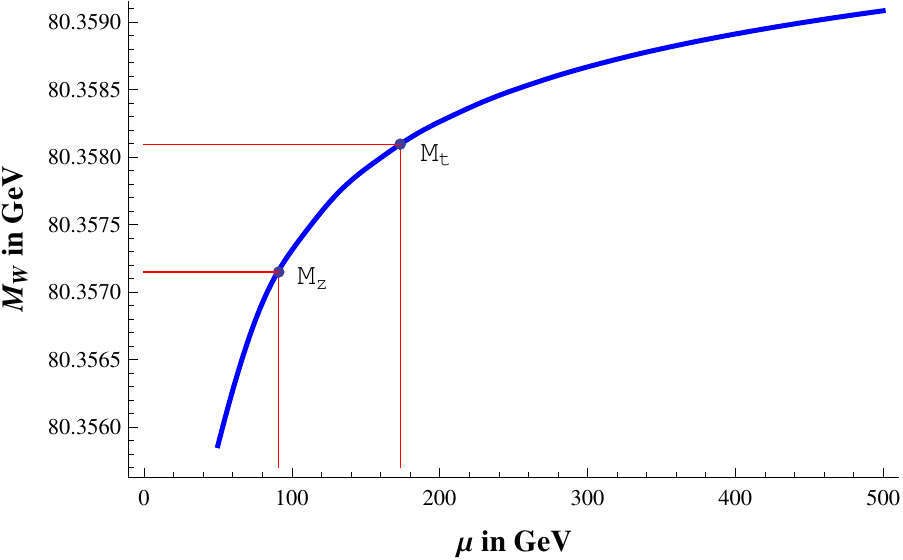}    
\caption{\label{fig1} Dependence of the $\mw$ prediction on the electroweak
scale $\mu$ in the \bms\ framework.}
\end{center}
\end{figure}

The result for the W mass described by eq.\,(\ref{parw}) is obtained
fixing $\mu = \mz$. As a physical quantity, the W mass must be 
$\mu$-independent. Hence the numerical difference between results obtained
varying $\mu$ in a ``reasonable'' interval can be taken as an indication of
the size of the missing higher-order corrections. In Fig.\,\ref{fig1}
we plot $\mw$ vs. $\mu$, with the 't-Hooft mass varying between 50 GeV
and 500 GeV. The figure is obtained using as input parameters the
central values in Table \ref{table1}. 
The figure shows a maximum variation of $\sim 3$ MeV in the entire range 
while in the 
restricted range $100 \le \mu \le 200$ GeV  we find a maximum variation of 
$\sim 1$ MeV.

\section{Discussion and conclusions}
In this paper we have discussed the $\mw-\mz$ interdependence in the SM, in the
\bms\ framework of the radiative corrections. We have evaluated the parameters
$\alc, \drcarw$ and $\rhoh$ at the full two-loop level augmented  by all the  
presently known three-loop strong, EW and mixed contributions 
and by the  four-loop strong corrections. We have presented our results via
simple formulas that parameterizes the results in  terms of $\mh, \, \mt,\
\as$ and the 5-flavor hadronic contribution to the vacuum polarization.

Our  calculation of the $W$ mass in the
\bms\ framework   automatically incorporates
the Dyson resummation of the lowest order large contributions, i.e.\
the mass singularity logarithms and the effects that scale as powers of the
top mass.  This partial inclusion of terms that are beyond 
the presently  computed effects in the loop expansion 
is a solid ground  to estimate in a realistic way
the size of the missing higher-order contributions in 
the $\mw$ computation.  The very weak
residual $\mu$-dependence shown in Fig.\,\ref{fig1} indicates that the
 uncertainty that can be assigned to our \bms\ result due to the
truncation of the perturbative series is expected to be at most $\sim 3$ MeV.

For what concerns  the parametric uncertainties,
 after the discovery of the Higgs
boson and the precise measurement  of its mass, the  most important 
experimental 
errors that affects the theoretical determination of  $\mw$  
are the ones on  $\mt$ and $\Dah$.  The left column
of Table \ref{table3} shows that the sensitivity of $\mw$ to
$\mt$  is  more than twice that to $\Dah$.
In  our calculation the top mass is  an on-shell quantity, i.e.\ a pole
mass,  and
in Table \ref{table1} we have identified it with  the average of the Tevatron, CMS and ATLAS 
measurements. However,  at the present level of precision of
the experimental determination ($\pm 0.76\, \gev$)
this identification can be disputed in two aspects. $i)$
The top pole mass has an intrinsic non-perturbative ambiguity
of   ${\cal O}(\Lambda_{QCD})$ due to infrared renormalon
effects. \ $ii)$  The top mass
parameter extracted by the experiments, which we call $\mt^{MC}$, is  obtained from the comparison  between the kinematical reconstruction
of the top quark decay products and the  Monte Carlo simulations of
the corresponding event. 
Therefore $ \mt^{MC}$ is a parameter sensitive to the on-shell region
of the top quark but it cannot  be directly identified  with  
$\mt$. The offset between $\mt$ and $\mt^{MC}$ is difficult to quantify,
and has recently been estimated of ${\cal O}( 0.3-0.5)\, \gev$ \cite{topmass}. 
In our numerics we have assigned  a $1\,\gev$ uncertainty to $\mt$.

The $\mw$ result obtained
using the central values in Table \ref{table1}, $\mw = 80.357$ GeV, agrees
within one and a half   standard deviations  with the present experimental world average,
$\mw = (80.385 \pm 0.015)$ GeV. 
 However, increasing the top mass and decreasing  $\Dah$ by 1$\sigma$, i.e.\
using $\mt = 174.34\, \gev$
and $\Dah =0.02717$,
we find $\mw = 80.370\, \gev$ which is much closer to
the experimental world average.  
It is interesting to note that the 
precise determination of the  top mass plays a
 very important role also  in the analysis of the stability of the SM Higgs 
potential up to the Planck scale. In order to get a closer 
agreement  between the computed $\mw$ and the experimental result we saw
that  large values of $\mt$ are favored, while  vacuum stability in the
SM requires  quite low values for the top mass, $\mt < 171.36 \pm 0.46 \, \gev$
\cite{Degrassi, Buttazzo}. Using $\mt = 171.36\, \gev$ and for the other
inputs the central values in Table \ref{table1} we find
$\mw = 80.345$ GeV, which differs from the experimental world average by more
than two standard deviations.

Our  \bms\  result for $\mw$ can be compared with
the prediction of $\mw$  in the OS scheme 
of Ref.\,\cite{ACFW} to study 
the scheme dependence of the $\mw$ predictions.
Both   calculations   include the complete 
two-loop electroweak contributions,  higher-order
QCD corrections of ${\cal O}(\alpha \as)$ and ${\cal O}(\alpha \as^2)$,   the
higher-order mixed EW-QCD corrections ${\cal O}(\alpha^2 \as \mt^4)$, and
purely EW   ${\cal O}(\alpha^3 \mt^6)$ corrections. In our result also
the four-loop contribution ${\cal O}(\alpha \as^3 \mt^2)$ is included, but
 we do not take it into account in the comparison with  Ref.\,\cite{ACFW}.
The   \bms\ and OS calculations  differ however in several aspects. While in the
\bms\ framework  we exploit
the possibility of resumming lowest-order contributions, no resummation is attempted 
in the OS calculation. Furthermore, our computation
refers directly to $\mw$, while in the calculation of Ref.\,\cite{ACFW} the 
quantity predicted is $\Mw$ (see sect.~\ref{outline}), which  is then 
translated to $\mw$ with
the introduction of a correction factor  containing the $W$ boson 
width. Because the latter is not very well known
(the experimental uncertainty is presently around $2 \%$), the theoretical
result for $\Gamma_W$ is employed, thus  introducing  an additional uncertainty
in the OS result  estimated to be 1-2 MeV \cite{FHWW}.

Since the $\mw$ determinations in the \bms\ and OS scheme
are equivalent at the two-loop level but differ 
by the partial inclusion of higher-order contributions, 
their numerical difference 
can be taken as a good  
estimate of  missing higher-order effects. 
Taking as inputs in our calculation those used in 
 \cite{ACFW}, i.e.\ 
$\mh = 100\, \gev$, $\mt= 174.3\, \gev$ $\Dah = 0.027572$ and $\as = 0.119$,
we find $\mw= 80.3749\, \gev$, which should be compared with the value $\mw = 80.3800$
reported in Ref.\,\cite{ACFW}. If instead we take the central values in Table
\ref{table1} as inputs in eq.\,(9) of Ref.\,\cite{ACFW}, we find an
OS result $\mw = 80.3639 \, \gev$ to be compared with an \bms\ result 
$\mw = 80.3578\, \gev$. These numbers indicates that the \bms\ determination
is always lower than the OS one, and shows  a larger  difference with the
present experimental world average. Furthermore,  the estimate
$\delta \mw^{th} \approx 4$ MeV of the theoretical uncertainty from
unknown higher-order corrections reported in Ref.\,\cite{ACFW}  seems to be
slightly  optimistic. 
A more realistic value  is probably 
$\delta \mw^{th} \approx 6$ MeV.

Another indication that 
$ \delta \mw^{th} \approx 4$ MeV is probably an
underestimate 
comes from our \bms\  calculation. 
In our hybrid scheme 
the masses that appear  in the one-loop contributions are identified with pole
masses, and the gauge couplings with \bms\ quantities. 
In this framework once
the one-loop contributions are written  as we did in the Appendix, the expressions of the two-loop corrections follow. However, their evaluation has some residual ambiguity,
because one can always re-express the $W$ mass as $\mz$ and $ \hat{c}$, or vice versa.
As we said, our choice to express $\mw$ in terms of $\mz$ 
in the two-loop contributions is the one that minimizes the $\mu$-dependence, but other choices
are allowed. Trying several possibilities, we found a variation  of $\delta \mw \approx 4$ MeV
in our \bms\ result; $\mw$  can be  almost 3 MeV below our default choice, further amplifying the difference
between 
the \bms\ and OS schemes.


We have seen that in our \bms\ calculation $\delta\mw^{th}\approx 3$\,MeV, while the scheme 
dependence observed in the comparison with the OS scheme is around 6\,MeV.
However, in our \bms\ computation we exploit at best all the
present available information through the automatic resummation of the known 
contributions. Moreover,  
we expect to have better control over the unknown  higher-order contributions
than  in the OS scheme because in \bms\  the effects
 related to $\delta \rho$ are not enhanced by the numerical 
factor $\cq/\sq$. Finally, we predict directly $\mw$ and not $\Mw$ in order to avoid  correction factors that introduce additional uncertainties.
It is therefore natural that the theoretical uncertainty estimated in the \bms\ 
calculation is smaller than the
scheme dependence in the comparison between the OS and \bms\ 
results. 


\subsection*{Acknowledgments}
The authors are indebted to A.~Freitas, M.~Gr\"unewald and
M.~Steinhauser for useful communications.  This work is supported in
part by MIUR under contract 2010YJ2NYW 006, by the EU Commission
under the Grant Agreement number PITN-GA-2010-264564 (LHCPhenoNet)
and the HiggsTools Initial Training Network PITN-GA-2012-316704,
and by Compagnia di San Paolo under contract ORTO11TPXK.
\begin{appendletterA}

\section*{Appendix}

Here we give the explicit formulae for $\Delta\hat{\alpha}$, $\drcarw$
and $Y_{\ms}$ at the one loop order. In the formulae below 
$N_c=3$ is the number of colors and
\be 
\mhw=\frac{\mh^2}{\mw^2}, \quad \mhz=\frac{\mh^2}{\mz^2}, \quad
\mbw=\frac{\mb^2}{\mw^2}, \quad\mbz=\frac{\mb^2}{\mz^2},  \quad
\mtw=\frac{\mt^2}{\mw^2},  \quad\mtz=\frac{\mt^2}{\mz^2}.
\ee  

\bea
\Delta\hat{\alpha}^{p,(1)}&=&-\frac{\alpha}{4\pi}\left\{\frac{2}{3}+\frac{4}{3}
\left(\lnb (m_e^2)+\lnb (m_{\mu}^2)+\lnb (m_{\tau}^2)\right)-7 \lnb (\mw^2)+
\frac{16}{27} N_c \lnb (\mt^2)\right.  \\
&&+\left. N_c \left(-\frac{196}{81}-\frac{4}{27}(1+2 \mbz)
B_0(\mz^2,\mb^2,\mb^2)-\frac{8}{27}\mbz\lnb(\mb^2)+
\frac{40}{27}\lnb(\mz^2)\right)\right\} \nonumber \label{Dalpha1}
\eea

\bea
\drcarwu&=&\frac{\hat{\alpha}}{4\pi \hat{s}^2} \left\{
\frac{1 + 8 \hat{c}^2}{12 c^4}
-\frac{7 + 80 \hat{c}^2}{24 \cq}
+\frac{1}{\hat{c}^2} - 8\hat{c}^2 + 
 \frac{1}{72} \left(794 - 21 \mhw + 6\mhw^2\right)
\right.\nonumber\\
&&+
\left(
\frac{\hat{c}^2 + 8 \hat{c}^4 + 64 \cq \hat{c}^4 - 6 c^4 (8 \hat{c}^4-5)}{12 \cq \hat{c}^2 \sq}
+\frac{26 \cq- \mhw^2 - 18 (5 + \mhz) + 
\mhw(82 + \mhz)}{12 \sq (1-\mhw)}
\right)
\lnb(\mw^2)\nonumber \\
&&+\left(
\frac{36 c^6+24 c^4+4 \cq-1}{12 c^4 \sq}
-\frac{3 \cq+2}{2 \hat{c}^2 \sq}
-\frac{\hat{c}^2 (15 c^4-11 \cq+2)}{3 c^4 \sq}
\right)
\lnb (\mz^2)\nonumber\\
&&+\frac{\mhw(12 - 4 \mhw + \mhw^2)}{12 (1-\mhw)} \lnb (\mh^2)
-\frac{12 - 4 \mhw + \mhw^2}{12} B_0(\mw^2,\mh^2,\mw^2)\nonumber\\
&&+\left(\frac{11 \hat{c}^2+1}{3\cq}+4 \hat{c}^2-\frac{8 \hat{c}^2+1}{12 c^4}-\frac{1}{\hat{c}^2}+2\right)B_0(\mw^2,\mz^2,\mw^2)\nonumber\\
&&+N_c \left[
\frac{1}{12} \left(4 \mbw \mtw-
2 \mbw^2-3\mbw-2 \mtw^2-3\mtw-12\right)
+\frac23 \lnb (\mw^2)
\nonumber\right.\\
&&+\frac{\mbw \left((\mbw-\mtw)^2+\mbw+2\mtw\right) }{6 (\mbw-\mtw)} \lnb (\mb^2)+\frac{\mtw \left((\mtw-\mbw)^2+\mtw+2\mbw\right)}{6(\mtw-\mbw)} \lnb (\mt^2) \nonumber\\  
&&\left.\left.+\frac{1}{6} \left((\mtw-\mbw)^2+\mtw+\mbw-2\right)
B_0(\mw^2,\mt^2,\mb^2)\right]\right\} 
\label{Drcur1}
\eea

\bea
Y_{\ms}^{(1)}&=&\frac{\hat{\alpha}}{4\pi \hat{s}^2} \left\{
\frac{1 + 8 \hat{c}^2}{12 \cq^2}+
\frac{175 - 416 \hat{c}^2 + 240 \hat{c}^4}{36 \cq}+
\frac{262 - 288 \hat{c}^2 + 3 \mhw^2 - 3 \mhw \mhz}{36}
\right.\nonumber\\
&&+	\left(
\frac{1 + 8 \hat{c}^2}{12 \cq}
+\frac{\mhw-30 + 64 \hat{c}^2 - 48\hat{c}^4}{12}
\right)\lnb (\mw^2)\nonumber \\
&&-\left(
\frac{1 + 8 \hat{c}^2}{12 c^4}+
\frac{\mhz+34 - 96 \hat{c}^2 + 48 \hat{c}^4}{12 \cq}
\right)\lnb (\mz^2)
-\frac{1}{12} \mhw^2 \sq \lnb (\mh^2)\nonumber\\
&&+\left(\frac{11 \hat{c}^2+1}{3 \cq}+
4 \hat{c}^2-\frac{8 \hat{c}^2+1}{12c^4}-\frac{1}{\hat{c}^2}+2\right)B_0(\mw^2,\mz^2,\mw^2)\nonumber\\
&&+\left(
\frac{1 - 4 \hat{c}^2 - 36 \hat{c}^4}{12 \cq}+
\frac{5- 8 \hat{c}^2 - 12 \hat{c}^4}{3}
\right)B_0(\mz^2,\mw^2,\mw^2)
\nonumber\\
&&-\left(\frac{(\mhw-4) \mhw}{12}+1\right)B_0(\mw^2,\mh^2,\mw^2)+
\left(\frac{(\mhz-4)\mhw}{12}+\frac{1}{\hat{c}^2}\right)B_0(\mz^2,\mh^2,\mz^2)\nonumber\\
&&+N_c \left[
\frac{11 - 22 \hat{c}^2 + 20\hat{c}^4}{9 \cq}
-\frac{(\mtw-\mbw)^2}{6}-1+\frac{2}{3}\lnb (\mw^2)-\frac{40 \hat{c}^4-44 \hat{c}^2+22}{27\cq} \lnb (\mz^2)
\right.\nonumber\\
&&+
\frac{9(\mbw-\mtw)-8 - 8 \hat{c}^2 + 16\hat{c}^4}{54}
 \mbw\lnb (\mb^2)
\nonumber\\
&&+
\frac{9(\mtw-\mbw)+16 - 80 \hat{c}^2 + 64\hat{c}^4}{54}
\mtw\lnb (\mt^2)\nonumber\\
&&+
\frac{5 - 4 \hat{c}^2 + 8 \hat{c}^4+
\left(16 \hat{c}^4- 8 \hat{c}^2-17\right)\mbz}{54 \cq}
B_0(\mz^2,\mb^2,\mb^2)\nonumber\\
&&+
\frac{17 - 40 \hat{c}^2 + 32 \hat{c}^4+
\left(64 \hat{c}^4- 80 \hat{c}^2+7\right)\mtz}{54 \cq} 
B_0(\mz^2,\mt^2,\mt^2)\nonumber\\
&&+\left.\left.\frac{(\mtw-\mbw)^2+\mbw+\mtw-2}{6} 
B_0(\mw^2,\mt^2,\mb^2)\right]\right\}
\label{Yms1l}
\eea
where $B_0$ is the 
finite part of the Passarino-Veltman function defined as
\be
B_0(s,x,y)=-\int_0^1 dt \lnb[tx+(1-t)y-t(1-t)s]
\label{B0}
\ee
where  $\lnb(x)=\log\left(\frac{x}{\mu}\right)$ with $\mu$ the energy scale.

\end{appendletterA}


\begin{thebibliography}{99}

\bibitem{higgsdiscovery}
Aad G.{\it et al.}  [ATLAS Collaboration],
  Phys.\ Lett.\ B {\bf 716} (2012) 1    [arXiv:1207.7214];
 Chatrchyan S.{\it et al.}  [CMS Collaboration],
  Phys.\ Lett.\ B {\bf 716} (2012) 30
  [arXiv:1207.7235].

\bibitem{Si81}
  A.~Sirlin and W.~J.~Marciano,
  Nucl.\ Phys.\ B {\bf 189} (1981) 442.


\bibitem{Si80}
  A.~Sirlin,
  Phys.\ Rev.\ D {\bf 22} (1980) 971.

\bibitem{Si84}
  A.~Sirlin,
  Phys.\ Rev.\ D {\bf 29} (1984) 89.

\bibitem{QCD2lrho}
  A.~Djouadi and C.~Verzegnassi,
  Phys.\ Lett.\ B {\bf 195} (1987) 265;


 \bibitem{QCD3lrho}
  L.~Avdeev, J.~Fleischer, S.~Mikhailov and O.~Tarasov,
  Phys.\ Lett.\ B {\bf 336} (1994) 560
   [Erratum-ibid.\ B {\bf 349} (1995) 597]
  [hep-ph/9406363];
 K.~G.~Chetyrkin, J.~H.~Kuhn and M.~Steinhauser,
  Phys.\ Lett.\ B {\bf 351} (1995) 331
  [hep-ph/9502291],

 \bibitem{QCD3lrhon}
 K.~G.~Chetyrkin, J.~H.~Kuhn and M.~Steinhauser,
  Phys.\ Rev.\ Lett.\  {\bf 75} (1995) 3394
  [hep-ph/9504413].





\bibitem{Hoo}
  J.~J.~van der Bij and F.~Hoogeveen,
  Nucl.\ Phys.\ B {\bf 283} (1987) 477.

\bibitem{bar}
  R.~Barbieri, M.~Beccaria, P.~Ciafaloni, G.~Curci and A.~Vicere,
  Phys.\ Lett.\ B {\bf 288} (1992) 95
   [Erratum-ibid.\ B {\bf 312} (1993) 511]
  [hep-ph/9205238]; 
  Nucl.\ Phys.\ B {\bf 409} (1993) 105.
  J.~Fleischer, O.~V.~Tarasov and F.~Jegerlehner,
  Phys.\ Lett.\ B {\bf 319} (1993) 249.

\bibitem{Con}
  M.~Consoli, W.~Hollik and F.~Jegerlehner,
  Phys.\ Lett.\ B {\bf 227} (1989) 167.


\bibitem{DGV}
  G.~Degrassi, P.~Gambino and A.~Vicini,
  Phys.\ Lett.\ B {\bf 383} (1996) 219
  [hep-ph/9603374].

\bibitem{DGS}
  G.~Degrassi, P.~Gambino and A.~Sirlin,
  Phys.\ Lett.\ B {\bf 394} (1997) 188
  [hep-ph/9611363].



\bibitem{FKSV}
  M.~Faisst, J.~H.~Kuhn, T.~Seidensticker and O.~Veretin,
  Nucl.\ Phys.\ B {\bf 665} (2003) 649
  [hep-ph/0302275];
 J.~J.~van der Bij, K.~G.~Chetyrkin, M.~Faisst, G.~Jikia and T.~Seidensticker,
  Phys.\ Lett.\ B {\bf 498} (2001) 156
  [hep-ph/0011373].

\bibitem{QCD2l}
 A.~Djouadi,
  Nuovo Cim.\ A {\bf 100} (1988) 357;
 B.~A.~Kniehl,
  Nucl.\ Phys.\ B {\bf 347} (1990) 86;
 A.~Djouadi and P.~Gambino,
  Phys.\ Rev.\ D {\bf 49} (1994) 3499
   [Erratum-ibid.\ D {\bf 53} (1996) 4111]
  [hep-ph/9309298].

\bibitem{FHWW}
  A.~Freitas, W.~Hollik, W.~Walter and G.~Weiglein,
  Phys.\ Lett.\ B {\bf 495} (2000) 338
   [Erratum-ibid.\ B {\bf 570} (2003) 260]
  [hep-ph/0007091];
  Nucl.\ Phys.\ B {\bf 632} (2002) 189
   [Erratum-ibid.\ B {\bf 666} (2003) 305]
  [hep-ph/0202131].

\bibitem{AC}
  M.~Awramik and M.~Czakon,
  Phys.\ Lett.\ B {\bf 568} (2003) 48
  [hep-ph/0305248].


\bibitem{ACOV}
  M.~Awramik and M.~Czakon,
  Phys.\ Rev.\ Lett.\  {\bf 89} (2002) 241801
  [hep-ph/0208113];
  A.~Onishchenko and O.~Veretin,
  Phys.\ Lett.\ B {\bf 551} (2003) 111
  [hep-ph/0209010];
 M.~Awramik, M.~Czakon, A.~Onishchenko and O.~Veretin,
  Phys.\ Rev.\ D {\bf 68} (2003) 053004
  [hep-ph/0209084].


\bibitem{ACFW}
  M.~Awramik, M.~Czakon, A.~Freitas and G.~Weiglein,
  Phys.\ Rev.\ D {\bf 69} (2004) 053006
  [hep-ph/0311148].

\bibitem{fitpro}
 D.~Y.~.Bardin, P.~Christova, M.~Jack, L.~Kalinovskaya, A.~Olchevski, S.~Riemann and T.~Riemann,
  Comput.\ Phys.\ Commun.\  {\bf 133} (2001) 229
  [hep-ph/9908433];
  H.~Flacher, M.~Goebel, J.~Haller, A.~Hocker, K.~Monig and J.~Stelzer,
  Eur.\ Phys.\ J.\ C {\bf 60} (2009) 543
   [Erratum-ibid.\ C {\bf 71} (2011) 1718]
  [arXiv:0811.0009 [hep-ph]];

\bibitem{Gfitter}
 M.~Baak, M.~Goebel, J.~Haller, A.~Hoecker, D.~Kennedy, R.~Kogler, K.~Moenig and M.~Schott {\it et al.},
  Eur.\ Phys.\ J.\ C {\bf 72} (2012) 2205
  [arXiv:1209.2716 [hep-ph]].

\bibitem{CFMS}
  M.~Ciuchini, E.~Franco, S.~Mishima and L.~Silvestrini,
  JHEP {\bf 1308} (2013) 106
  [arXiv:1306.4644 [hep-ph]].


\bibitem{Si89}
  A.~Sirlin,
  Phys.\ Lett.\ B {\bf 232} (1989) 123.

\bibitem{FS}
 S.~Fanchiotti and A.~Sirlin,
  Phys.\ Rev.\ D {\bf 41} (1990) 319.

\bibitem{DFS}
  G.~Degrassi, S.~Fanchiotti and A.~Sirlin,
  Nucl.\ Phys.\ B {\bf 351} (1991) 49.

\bibitem{DV}
  G.~Degrassi and A.~Vicini,
  Phys.\ Rev.\ D {\bf 69} (2004) 073007
  [hep-ph/0307122].

\bibitem{Si91}
  A.~Sirlin,
  Phys.\ Rev.\ Lett.\  {\bf 67} (1991) 2127;
  S.~Willenbrock and G.~Valencia,
  Phys.\ Lett.\ B {\bf 259} (1991) 373.

\bibitem{BLRS}
  D.~Y.~Bardin, A.~Leike, T.~Riemann and M.~Sachwitz,
  Phys.\ Lett.\ B {\bf 206} (1988) 539.

\bibitem{PDG}
  K.~A.~Olive {\it et al.}  [Particle Data Group Collaboration],
  Chin.\ Phys.\ C {\bf 38} (2014) 090001.

\bibitem{Feynart}
  T.~Hahn,
  Comput.\ Phys.\ Commun.\  {140} (2001) 418
  [hep-ph/0012260].

\bibitem{Mertig:1998vk}
  R.~Mertig and R.~Scharf,
  Comput.\ Phys.\ Commun.\  {111} (1998) 265
 [hep-ph/9801383].

\bibitem{Tarasov}
  O.~V.~Tarasov,
  Nucl.\ Phys.\ B {502} (1197) 455
  [hep-ph/9703319].


\bibitem{Feyncalc}
  R.~Mertig, M.~Bohm and A.~Denner,
  Comput.\ Phys.\ Commun.\  {64} (1991) 345.

\bibitem{DT}
  A.~I.~Davydychev and J.~B.~Tausk,
  Nucl.\ Phys.\ B {\bf 397} (1993) 123.

  
\bibitem{Martinint}
  S.~P.~Martin,
  Phys.\ Rev.\ D {68} (2003) 075002
 [hep-ph/0307101].


\bibitem{MartinTSIL}
  S.~P.~Martin and D.~G.~Robertson
    Comput.\ Phys.\ Commun.\  {174} (2006) 133
  [hep-ph/0501132].
%


\bibitem{DHMZ}
  M.~Davier, A.~Hoecker, B.~Malaescu and Z.~Zhang,
  Eur.\ Phys.\ J.\ C {\bf 71} (2011) 1515
   [Erratum-ibid.\ C {\bf 72} (2012) 1874]
  [arXiv:1010.4180 [hep-ph]].

\bibitem{BP}
  H.~Burkhardt and B.~Pietrzyk,
  Phys.\ Rev.\ D {\bf 84} (2011) 037502
  [arXiv:1106.2991 [hep-ex]].


\bibitem{QCD3la}
 K.~G.~Chetyrkin, J.~H.~Kuhn and M.~Steinhauser,
  Nucl.\ Phys.\ B {\bf 482} (1996) 213
  [hep-ph/9606230].


\bibitem{Kinoshita:1958ru}
  T.~Kinoshita and A.~Sirlin,
  Phys.\ Rev.\  {113} (1959) 1652.


\bibitem{Deltaq}    T.~van Ritbergen and R.~G.~Stuart,
  Nucl.\ Phys.\ B {564} (2000) 343
  [hep-ph/9904240]. 

\bibitem{Buttazzo}
  D.~Buttazzo, G.~Degrassi, P.~P.~Giardino, G.~F.~Giudice, F.~Sala, A.~Salvio and A.~Strumia,
  JHEP {\bf 1312} (2013) 089
  [arXiv:1307.3536 [hep-ph]].

\bibitem{topmass}
 A.~H.~Hoang and I.~W.~Stewart,
  Nucl.\ Phys.\ Proc.\ Suppl.\  {\bf 185}, 220 (2008)
  [arXiv:0808.0222 [hep-ph]];\\
 S.~Moch, S.~Weinzierl, S.~Alekhin, J.~Blumlein, L.~de la Cruz, 
 S.~Dittmaier, M.~Dowling and J.~Erler {\it et al.},
  arXiv:1405.4781 [hep-ph].

\bibitem{Degrassi}
  G.~Degrassi, S.~Di Vita, J.~Elias-Miro, J.~R.~Espinosa, G.~F.~Giudice, G.~Isidori and A.~Strumia,
  JHEP {\bf 1208} (2012) 098
  [arXiv:1205.6497 [hep-ph]].

\end{thebibliography}
\end{document}